\documentclass{article}

\usepackage{amsmath}
\usepackage{PRIMEarxiv}
\usepackage[utf8]{inputenc} % allow utf-8 input
\usepackage[T1]{fontenc}    % use 8-bit T1 fonts
\usepackage{hyperref}       % hyperlinks
\usepackage{url}            % simple URL typesetting
\usepackage{booktabs}       % professional-quality tables
\usepackage{amsfonts}       % blackboard math symbols
\usepackage{nicefrac}       % compact symbols for 1/2, etc.
\usepackage{microtype}      % microtypography
\usepackage{lipsum}
\usepackage{fancyhdr}       % header
\usepackage{subfigure}       % figure
\usepackage{graphicx}       % graphics
\graphicspath{ {./images/}} 
\usepackage{natbib}
\usepackage{multirow}

\makeatletter
\DeclareRobustCommand{\cev}[1]{%
  {\mathpalette\do@cev{#1}}%
}
\newcommand{\do@cev}[2]{%
  \vbox{\offinterlineskip
    \sbox\z@{$\m@th#1 x$}%
    \ialign{##\cr
      \hidewidth\reflectbox{$\m@th#1\vec{}\mkern4mu$}\hidewidth\cr
      \noalign{\kern-\ht\z@}
      $\m@th#1#2$\cr
    }%
  }%
}
\makeatother

\newcommand*{\affaddr}[1]{#1} % No op here. Customize it for different styles.
\newcommand*{\affmark}[1][*]{\textsuperscript{#1}}
\newcommand*{\email}[1]{\texttt{#1}}

\begin{document}
  
%%%%%%%%%%%%%%% Title ===> insert title %%%%%%%%%%%%%%%%%%%%%%%%%
\title{ Finding essential parts of the brain in rs-fMRI can improve diagnosing ADHD by Deep Learning
%%%% Cite as
%%%% Update your official citation here when published 
%\thanks{\textit{\underline{Citation}}: 
%\textbf{Authors. Title. Pages.... DOI:000000/11111.}} 
}

\author{
Byunggun Kim\affmark[1], Jaeseon Park\affmark[1], Taehun Kim \affmark[1], and Younghun Kwon\affmark[1,2]\\
\affaddr{\affmark[1]Department of Applied Artificial Intelligence}\\
\affaddr{\affmark[1]Department of Applied Physics}\\
\email{\{byunggunkim, wotjs307, taehunkim, yyhkwon\}@hanyang.ac.kr}\\
\affaddr{Hanyang University, Ansan, Kyunggi-Do, 425-791, Republic of Korea}%
}

%\author{
%  %% First Author
%  Byunggun Kim \\
%  Hanyang University \\
%  Ansan\\
%  \texttt{byunggunkim@hanyang.ac.kr} \\
%  %% examples of more authors
%   \And
%     Jaeseon Park \\
%  Hanyang University \\
%  Ansan\\
%  \texttt{wotjs307@hanyang.ac.kr} \\
%  \And
%  Taehun Kim \\
%  Hanyang University \\
%  Ansan\\
%  \texttt{taehunkim@hanyang.ac.kr} \\
%  \And
%  Younghun Kwon \\
%  Hanyang University \\
%  Ansan\\
%  \texttt{yyhkwon@hanyang.ac.kr} \\
%    %% \AND
%  %% Coauthor \\
%  %% Affiliation \\
%  %% Address \\
%  %% \texttt{email} \\
%  %% \And
%  %% Coauthor \\
%  %% Affiliation \\
%  %% Address \\
%  %% \texttt{email} \\
%  %% \And
%  %% Coauthor \\
%  %% Affiliation \\
%  %% Address \\
%  %% \texttt{email} \\
%}

\maketitle

%%%%%%%%%%%%%%%%% abstract %%%%%%%%%%%%%%%%%%%%%%%
\begin{abstract}
Attention-Deficit/Hyperactivity Disorder(ADHD) is considered a very common psychiatric disorder, but it is difficult to establish an accurate diagnostic method for ADHD. Recently, with the development of computing resources and machine learning methods, studies have been conducted to classify ADHD using resting-state functional magnetic resonance(rs-fMRI) imaging data. However, most of them utilized all areas of the brain for training the models. In this study, as a different way from this approach, we conducted a study to classify ADHD by selecting areas that are essential for using a deep learning model. For the experiment, rs-fMRI data provided by ADHD-200 global competition was used.
To obtain an integrated result from the multiple sites, each region of the brain was evaluated with ‘Leave-one-site-out’ cross-validation. As a result, when we only used 15 important region of interest(ROIs) for training, an accuracy of 70.6\% was obtained, significantly exceeding the existing results of 68.6\% from all ROIs. In addition, to explore the new structure based on SCCNN-RNN, we performed the same experiment with three modified models: (1) Separate Channel CNN - RNN with Attention (ASCRNN), (2) Separate Channel dilate CNN - RNN with Attention (ASDRNN), (3) Separate Channel CNN - slicing RNN with Attention (ASSRNN). As a result, the ASSRNN model provided a high accuracy of 70.46\% when training with only 13 important region of interest (ROI). These results show that finding and using the crucial parts of the brain in diagnosing ADHD by Deep Learning can get better results than using all areas.
\end{abstract}

% keywords can be removed

%%%%%%%%%%%%%%%%% introduction %%%%%%%%%%%%%%%%%%%%%%%
\section{Introduction}
\label{sec:introduction}
Attention-Deficit/Hyperactivity Disorder(ADHD) is known as a psychiatric disorder that frequently appears in children. (\cite{barkley1997behavioral, faraone2003worldwide, polanczyk2007worldwide}) However, an accurate diagnostic method for ADHD is not known yet. (\cite{barkley1997behavioral}) To overcome this difficulty, there have been efforts to find a biomarker between healty control(HC) and ADHD. For example, functional connectivity (FC) is extracted from fMRI data based on machine learning methods. (\cite{RN153, RN146, RN151, guo2014adhd, RN152, RN148, RN149}) They tried to classify ADHD using FC.

In recent years, the models with deep neural networks (\cite{riaz2017fcnet, RN1, riaz2018deep, RN8, RN7})  are also used to get the feature. These methods obtained high accuracies compared to the traditional methods. Most of the studies mentioned so far used the whole regions of the brain.

However, we need to focus on the results that the difference feature between ADHD and HC might be in a specific or some of the region of interest (ROI). (\cite{RN144, RN152}) Therefore, it is natural that we have a question ‘Is it good to cover all brain areas in the diagnosis of ADHD through a neural network model?’

In this study, we want to find a specific answer to the question. The chosen dataset, evaluation method, and models are as follows. At first, because the measure parameter was different with each site, we combined all sites of data as the training dataset to avoid biased results. Next, we evaluated the trained model with Leave-One-Site-Out Cross-validation(LOSO) (\cite{RN7}) and compared it, using the ROIs that we used for the training. The model architectures for this experiment are ‘Separated channel CNN - RNN(SCCNN-RNN)’ proposed by  (\cite{RN7}) and the new architecture based on it. Each of the models has the same number of trainable parameters regardless of the number of ROIs. It means that we can control the variance of the result from the model capacity. So we can get a more accurate result related to the importance of the ROIs.

The experimental procedures were also designed to examine the existence of some critical areas in the identification of ADHD. First of all, with the SCCNN-RNN model, we examine the importance of the individual ROI by using only one ROI feature for the model training. Then, the ROIs were ranked according to the results. In the second experiment, By selecting some of the ranked ROI that have a considerable influence on the diagnosis of ADHD, we investigated how it affects the classification accuracy. We found in the second experiment that using only high-ranked ROIs is much better than using whole ROIs for the classification. We conducted experiments on three other new architecture models based on ‘SCCNN-RNN’ to supplement this result.

In conclusion, even with a small amount of ROIs, the evaluation of the result show 70.6\% accuracy. In the next \ref{sec:method}, \ref{sec:experiment} sections, we will explain the data selection and descriptions of the models for the experiment.

%%%%%%%%%%%%%%%%% Method %%%%%%%%%%%%%%%%%%%%%%%
\section{Method}
\label{sec:method}

%%%%%%%%%%%%%%%%% Method subsection1 %%%%%%%%%%%%%%%%%%%%%%%
\subsection{To diagnosis the ADHD, is it necessary that we take the whole regions of the brain?}
\label{sec:method1}

 As mentioned in the Introduction, several studies have been conducted to find the biomarker between HC and ADHD patients. In recent years, the models with deep neural networks have been used to understand the biomarker. In order to do it, they used the whole regions of the brain. 

However, there were the results that the difference feature between ADHD and HC might be in a specific or some of the ROIs. (\cite{RN144, RN152})  Along this line,  we conduct a study to answer the question:Is it good to cover all brain areas in the diagnosis of ADHD through a neural network model?

Fortunately, in our study, we may answer the question. As the result of our investigation, we can show that the SCCNN-RNN model using a small amount of ROIs provides 70.6\% accuracy, which exceeds the existing results of 68.6\% using all ROIs.

%%%%%%%%%%%%%%%%% Method subsection2 %%%%%%%%%%%%%%%%%%%%%%%
\subsection{Data selection for experiment (Why we use the AAL 116 template)}
\label{sec:method2}

The rs-fMRI, as the 4-dimensional structure, has both spatial and temporal information of the brain. Therefore, the raw data obtained from a single subject contains a large amount of low-dimensional (x, y, z, t) features. It means that we need lots of data samples to learn a meaningful hidden feature of ADHD with a neural network. Fortunately, ‘Neuro Bureau’ (\cite{RN9}) provides many data samples($\sim$1k) preprocessed with various methods (Athena, NIAK, Burner) used in the ADHD-200 Global competition. Nevertheless, there are some difficulties in using them directly for the experiment. For the first reason, as we said before, The number of data samples is still small enough to use low-dimensional data for training directly.

Furthermore, the second reason is ADHD-200 Global competition’s dataset consisted of several sites. In other words, Each site of fMRI data was collected with different parameters of MRI devices. If one trains the model with the fMRI data samples from the specific site, it occurs biased result to the initial setting of the measuring device. 

In this study, we try to use as many data samples as possible and obtain results that are not dependent on the measuring device. To overcome these situations, we constructed a training dataset from multiple sites together. And we extracted feature vectors that are less sensitive to the unique biological information (phenotype) and measure parameters. In previous studies, handcraft feature extraction (\cite{RN17, RN18}) was frequently used. However, these can depend on the context in which the fMRI data are measured. So it is not a proper method for our situations. For that reason, we take the Automated Anatomical Labelling(AAL 116) (\cite{RN145}) for the feature extraction method.

If we use the AAL 116 template, we expect it is possible to effectively extract features, according to contribution on the 116 interest regions (ROIs) in low-dimensional fMRI data. In summary, we choose the dataset from the five sites (NYU, Peking, OHSU, KKI, NI) preprocessed by the NIAK pipeline(\cite{lavoie2012integration}) from The Neuro Bureau ADHD-200 preprocessed Repository. And then, we extract high-level features using the AAL template. Through these processes, we showed that a specific ROI is useful for diagnosing ADHD in general fMRI data.

%%%%%%%%%%%%%%%%% Method subsection3 %%%%%%%%%%%%%%%%%%%%%%%
\subsection{Separate Channel CNN - RNN Architecture}
\label{sec:method3}

%% figure sccnn-rnn structure %%
\begin{figure}[ht]
  \centering
  \includegraphics[scale=0.85]{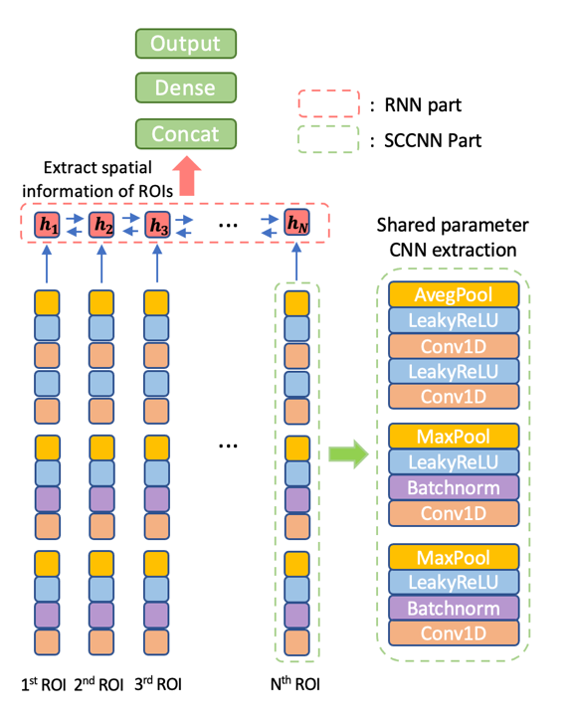}
  \caption{Architecture of SCCNN - RNN. After passing the SCCNN part that extracts the bold signal of each ROI through convolution layers and the RNN part that learns the relationship between multiple ROIs, ADHD is determined from the output of the last step of RNN through two fully connected layers.}
  \label{fig:fig1}
\end{figure}

As we mentioned before, we want to find out the meaningful ROIs for diagnosing ADHD and HCs. To this end, we selected ‘the Separate Channel CNN - RNN(SCCNN-RNN)’ proposed by (\cite{RN7}) as the base model architecture for the experiment, which satisfies the following two reasons.

The first reason is that ‘SCCNN-RNN’ can extract spatial and temporal information from fMRI data. Specifically, ‘SCCNN-RNN’ can be divided into two parts with different purposes. The SCCNN part can extract the feature of the BOLD signal in each ROI with 1-D CNN. The RNN part can learn the spatial relation of the ROIs.

And the second reason is that ‘SCCNN-RNN’ architecture can always keep the same number of learnable weights regardless of the change in the input data dimension. For example, we train several models with different ROIs and compare them with evaluation results. If the input’s shape changes the number of trainable parameters, the model’s learning capacity can also be changed. It means that we compared the influence of the ROI with inconsistent results. Therefore, to avoid this situation, we controlled the trainable parameters.

Now, we explain a detailed setting of the SCCNN-RNN for our experiments. It is shown in Fig \ref{fig:fig1}. In the SC-CNN part, we stack four layers with 1D CNN. The convolution layer’s channel number is ‘32’, ‘64’, ‘96’, ‘96’, respectively. And the stride size is ‘1’, and the filter size is ‘3’ as common parameters on the convolution layers. In the RNN part, we used the Bidirectional LSTM cell(\cite{graves2005framewise}) because it's mechanisms were proved the performance in many sequence data domains such as speech recognition (\cite{graves2013speech}) and language model (\cite{sundermeyer2012lstm}) and so on. The hidden state numbers of each step are set to ‘128’. The output of the last T-step of the RNN are connected with one fully connected layer with ‘128’ neurons. At last, the classification layer is put as the last layer with a softmax activation to output the two-dimension vectors as the probability of ADHD and HCs.

As we said, the ‘SCCNN-RNN’ structure is a proper and straightforward model to learn the spatial and temporal features from the rs-fMRI data for our experiment. Therefore, in the next section, we expand this structure to explore the more efficient result.

%%%%%%%%%%%%%%%%% Method subsection4 %%%%%%%%%%%%%%%%%%%%%%%
\subsection{Other modified models based on the ‘SCCNN - RNN’ architecture}
\label{sec:method4}

%% figure modified sccnn-rnn structure %%
\begin{figure}[t]
\centering
\subfigure[]{
\includegraphics[width=0.28\linewidth]{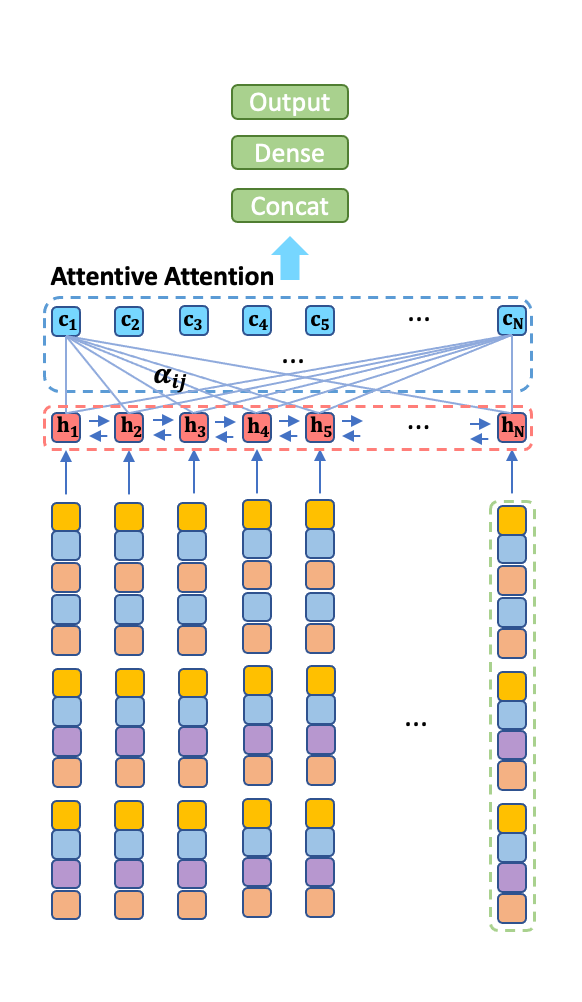}
\label{fig:fig2_a}
}
\centering
\subfigure[]{
\includegraphics[width=0.3\linewidth]{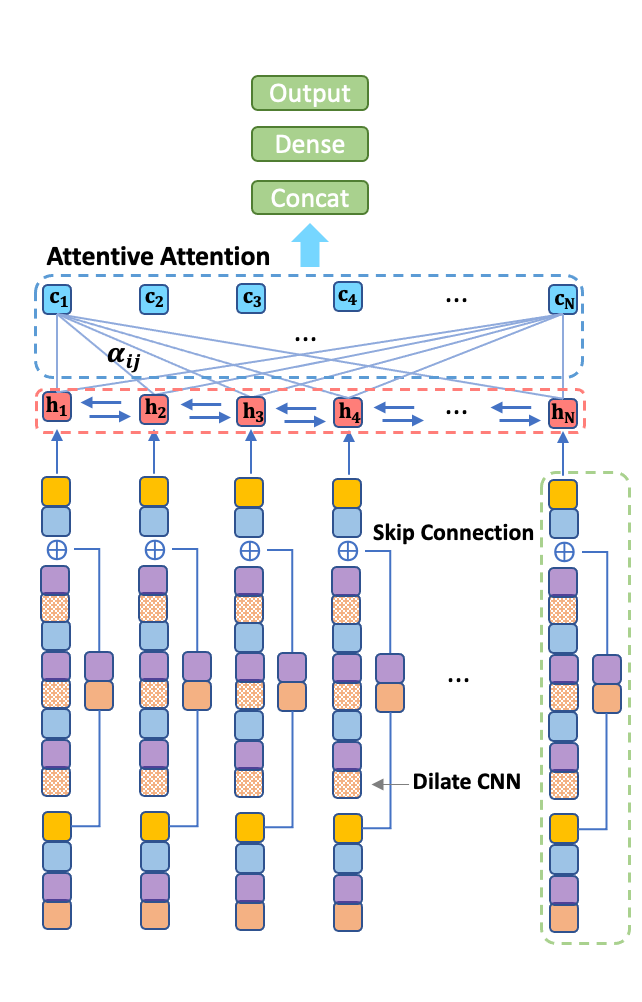}
\label{fig:fig2_b}
}
\centering
\subfigure[]{
\includegraphics[width=0.3\linewidth]{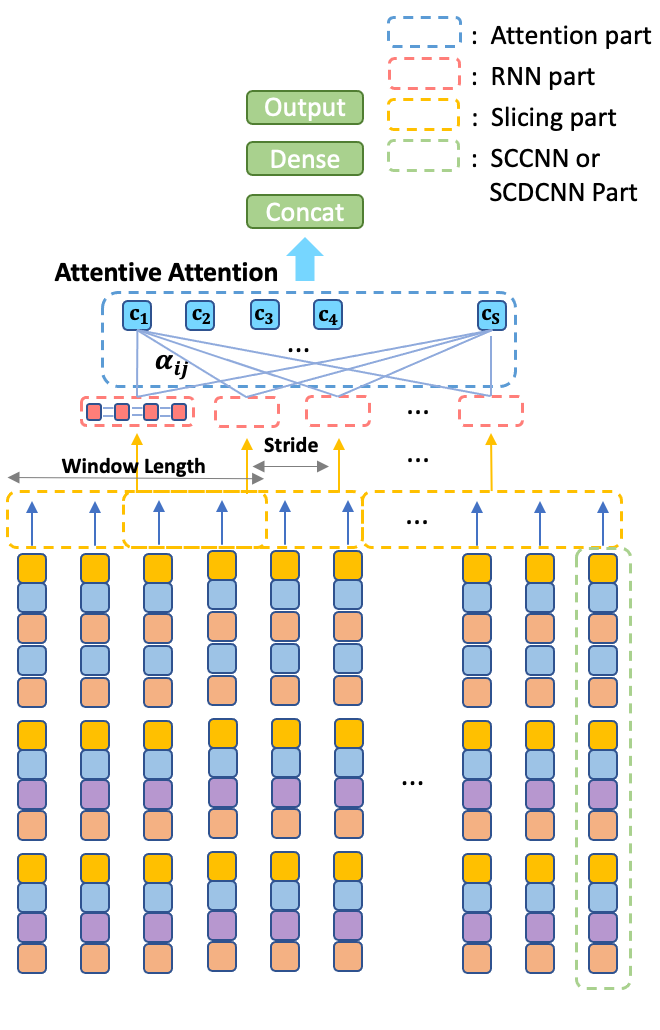}
\label{fig:fig2_c}
}
    \caption{Differences in the proposed model architectures. (a) Separate Channel CNN - RNN with Attention (ASCRNN) (b) Separate Channel dilate CNN - RNN with Attention (ASDRNN) (c) Separate Channel CNN - slicing RNN with Attention(ASSRNN). All models contain the attention mechanism ($c_i =\sum_{j=1}^{N_R} \alpha_{ij} h_j $$\colon$ attention weight, $h_j \colon$ output  in  $j$ step on the RNN, $R_N \colon$ Number of ROIs used in training) in common. In (a), only attention mechanism from SCCNN - RNN structure is applied. In (b), the Dilate CNN and skip connection on the SCCNN part based on the ASCRNN is applied. In (c), the slicing BiLSTM based on the ASCRNN is applied.}
\label{fig:fig2}
\end{figure}

To make the modified models from the ‘SCCNN - RNN’ structure, we applied some ideas from the speech emotion recognition(SER) model (\cite{meng2019speech, xie2019speech, peng2020speech}). In other words, we chose ideas of the three different speech emotion recognition models. The reason is that speech emotion recognition using a neural network also requires a structure for obtaining emotion information in a long sequence with a small amount of data. It is similar to our problems. For this reason, this study was conducted to investigate the neural network structure to obtain features from fMRI data and various SCCNN-RNN-based model structures. And it also shows that selecting and using a region with high impact is more helpful in disease identification than using all regions. 

 As you can see in Fig \ref{fig:fig2}, these three modified models from ‘SCCNN-RNN’ have slight differences. However, the attention mechanism is applied in common. The attention mechanism usually improves the performance of the models if the training data is sequential. In ‘SCCNN-RNN’ structure, it only uses the last hidden state outputs from BiLSTM as the next layer’s inputs. Therefore, it is structurally difficult to learn the importance of ROIs, To overcome it, we put into the attention mechanism to focus on the important ROI with all step’s hidden states.
 
 In the recent years, the attention mechanism has been used as several ways(\cite{bahdanau2014neural, luong2015effective, vaswani2017attention}). In our study, we chose attentive attention method proposed in (\cite{RN7}). Specifically, to learn the correlation between the two reference ROIs, after linear transformation of comparison vector with learnable weight matrix , map into non-linear function is considered. It should be noted that in this study, despite using the same method as the previous, there is difference. In (\cite{RN7}), the attention was stacked after ‘SCCNN’ part. but we stack in order of SCCNN part, RNN part, and attention mechanism. The reason for applying the attention method in this way is to consider not only the relationship between two areas, but also the relationship between several areas. For example, $h_j$ of jth BiLSTM’s hidden state consists of the forward hidden state $\vec{h}_j$ ,contained with information from $1$st ROI to $j$th ROI, and backward hidden state $\cev{h}_j$, contained with information from $R_N$ th ROI to $j$th ROI in reverse order. And we learn the correlation $\alpha_{ij}$ between $h_i$ and $h_j$ with the attention layer. Therefore all ROIs that we choose can be considered in one step. The equations \ref{eq:eq1}, \ref{eq:eq2} and \ref{eq:eq3} describe the process. We call this model as ‘Separate Channel CNN - RNN with Attention’ (ASCRNN). This attention is also applied to the two models proposed later in the same way. 

\begin{gather}
 h_j =  \left[\vec{h}_j,  \cev{h}_j \right], \quad 1\le j \le N_R,  \label{eq:eq1}\\
 \alpha_{ij} = attentive\left( h_i , h_j \right)  \label{eq:eq2} \\
c_i =\sum_{j=1}^{N_R} \alpha_{ij} h_j  \label{eq:eq3}
\end{gather}

\begin{figure}[ht]
  \centering
  \includegraphics{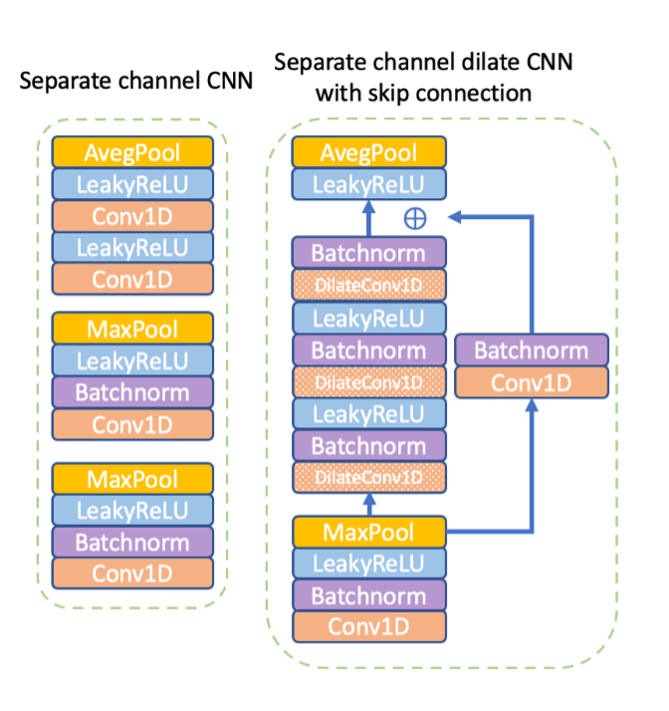}
  \caption{The comparison of separate channel CNN and Separate channel dilate CNN with skip connection architectures.}
  \label{fig:fig3}
\end{figure}

The next modified model called ‘Separate Channel dilate CNN - RNN with Attention(ASDRNN)’ is focused on the BOLD signal extraction. It differs from the SCCNN part. From previous studies (\cite{RN8}), it can be seen that learning with only the relationship within a specific frame of the BOLD signal is helpful in the diagnosis of ADHD. With this fact, we replace the 1-D CNN with dilation 1-D CNN (\cite{yu2015multi}). Also, to treat the gradient vanishing problem that emerged from the deep neural networks, we applied the skip connection (\cite{he2016deep}) after the last batch normalization layer. The detailed structure of Separate Channel dilated CNN can be seen in Fig \ref{fig:fig2_b} and \ref{fig:fig3}. In dilation 1-D CNN, dilate rate is set as ‘2’.

Fig \ref{fig:fig2_c} is the last modified model named Separate Channel CNN - slicing RNN with Attention(ASSRNN). This model takes a segmented area as input instead of giving the entire ROI area as input to the RNN to get the relationship between partitioned regions. To do that, instead of BiLSTM in the ASCRNN model, we applied the slicing BiLSTM(\cite{peng2020speech}). The slicing BiLSTM’s inputs are the series of subsequences splatted from the original sequence with constant window size$(w)$ and stride size$(l)$. This can be expected to focus on the relationship within the small number of ROIs and avoid the gradient vanishing problem that may appear as the time step of the input becomes longer in the RNN structure. This process can be found in below \ref{eq:eq4}, \ref{eq:eq5} and \ref{eq:eq6} equations

\begin{gather}
s = \left\{ 0, 1, ... ,   \lceil {N_R - l \over w} \rceil \right\} \label{eq:eq4} \\
x^{(s)} = \left\{ \begin{array}{rcl}
\left[ x_{1+sw}, ..., x_{l+sw} \right] & \mbox{if} & 0 \le s \le  \lceil {N_R - l \over w} \rceil -1 \\
\left[ x_{N_R -l + 1}, ..., x_{N_R}\right] & \mbox{else} & 
 \end{array}\right. \label{eq:eq5}\\
 h_s = BiLSTM \left( x^{(s)} \right)\label{eq:eq6}
\end{gather}

%%%%%%%%%%%%%%%%% experiment section %%%%%%%%%%%%%%%%%%%%%%%
\section{Experiment Result and Discussion}
\label{sec:experiment}

%%%%%%%%%%%%%%%%% experiment section1 %%%%%%%%%%%%%%%%%%%%%%%
\subsection{Settings in Training and Evaluation}
To analysis the result to choice of the ROIs, all parameters required for learning were chosen as identical ones in our experiments. The parameter setting is based on the results obtained through the experiment. Specifically, we used Adam optimizer (\cite{kingma2014adam}). The learning rate was chosen to be 1e-4. Xavier initialization(\cite{glorot2010understanding}) was used as the Initialization method for all trainable weights. Also, to prevent the overfitting problem, we used l2 regularization with factor 0.0005. The leaky ReLU (\cite{maas2013rectifier}) with 0.1 slope coefficient was chosen as the activation function. Next, to avoid bias due to an imbalanced dataset corresponding to ADHD and HC, the same number of each class (ADHD, HC) was sampled for all mini-batch. The mini-batch size was set to be 32. And then, we set binary cross entropy for the loss function. The evaluation method of the accuracy of the models was chosen by ‘Leave-one-site-out cross validation(LOSO)’. By proceeding with the evaluation with the data set of another site not used for training the model, it is possible to avoid the dependent characteristics (parameters of the measurement device). And the experiment results for the ROIs that play an essential role in diagnosis can be general.

%%%%%%%%%%%%%%%%% experiment section2 %%%%%%%%%%%%%%%%%%%%%%%
\subsection{Detail of the dataset composition}
We used rs-fMRI data samples provided by the ADHD-200 competition (\cite{RN9}) for the experiment. They consist of the data from eight institutions (NYU : New York University child study center, Peking : Peking University, OHSU : Oregon Health Sciences University, KKI : Kennedy Krieger Institute, NI : NeuroIMAGE, BHBU : Bradley/ Brown University, Pitt : University of Pittsburgh, WUSTL : Washington University at Saint Louis) Among them, BHBU, Pitt, and WUSTL are difficult to use for supervised learning, so five other sites (NYU, Peking, OHSU, KKI, NI) are used. Moreover, to compensate for the lack of data, all data (training data, test data) of each site were combined and considered one site dataset. So, the overall structure of the data we used is shown in Table \ref{tab:table1}. And rs-fMRI sample is preprocessed, as we mentioned before. See in section \ref{sec:method2}.

\begin{table}
  \centering
  \begin{tabular}{ccccccc}
    \toprule  
    & NYU & Peking & OHSU & KKI &  NI & Total \\
   \midrule
   ADHD & 147 & 101 & 43 & 25 &  36 & 352 \\
   HC & 110 & 143 & 70 & 69 &  37 & 429 \\
 Total  & 257 & 244 & 133 & 94 &  73 & 781 \\

   \bottomrule
  \end{tabular}
   \caption{The data composition of each site that we used in our experiments}
  \label{tab:table1}
\end{table}

%%%%%%%%%%%%%%%%% experiment section3 %%%%%%%%%%%%%%%%%%%%%%%
\subsection{Analysis selecting an important region of the brain according to AAL 116 ROI using a neural network model of SCCNN-RNN}
\label{sec:experiment3}
\begin{figure}[ht]
  \centering
  \includegraphics[scale=0.5]{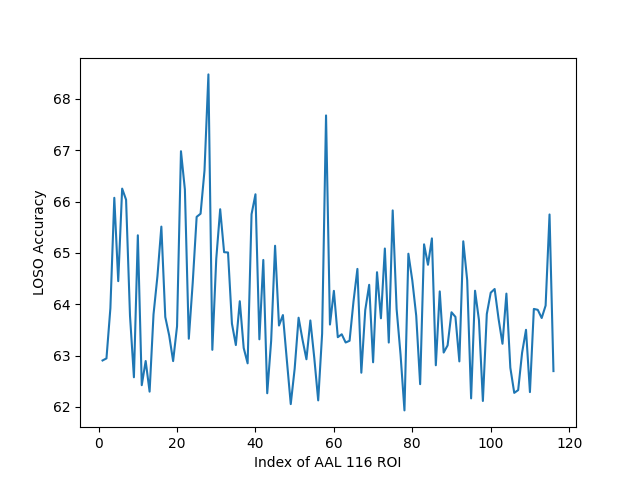}
  \caption{The accuracy according to individual ROI when we trained the SCCNN-RNN with only one individual ROI.}
  \label{fig:fig4}
\end{figure}

To determine the importance of individual ROIs for ADHD discrimination, we evaluated the accuracy through independent neural network model trained with specific ROI. For the experiment, we used ‘SCCNN-RNN’ models with LSTM. Fig \ref{fig:fig4} shows the diagnosis accuracy trained with specific ROI. The distribution of the accuracy obtained in the experiment is in a large range(minimum accuracy : 61.93\% with index=77, minimum accuracy : 68.47\% with index=28). This indicates a meaningful result when learning with only one ROI because it has fewer fMRI features than all ROIs. It can be expected that there may be a region that plays an important role in discriminating ADHD. Based on these results, in next experiment, we will show how learning by selecting a few important ROIs differs from using all ROIs.
 
%%%%%%%%%%%%%%%%% experiment section4 %%%%%%%%%%%%%%%%%%%%%%%
\subsection{Accuracy analysis for ADHD discrimination according to ROI ranking in SCCNN-RNN}
\label{sec:experiment4}
The experiment is performed to determine how the association between important ROIs can be understood in the ADHD discrimination. The specific progress of the experiment is as follows. According to the ROI ranking from the result of section \ref{sec:experiment3} , the regions at the top are selected sequentially, and the number of ROI features is gradually increased. And the experiment is carried out with the independent neural network models. Also, the same model(SCCNN-RNN)  is used in order to reduce the difference that will appear due to the change in the number of parameters of the neural network model according to the increase of the input’s data size.

\begin{figure}[ht]
  \centering
  \includegraphics[scale=0.5]{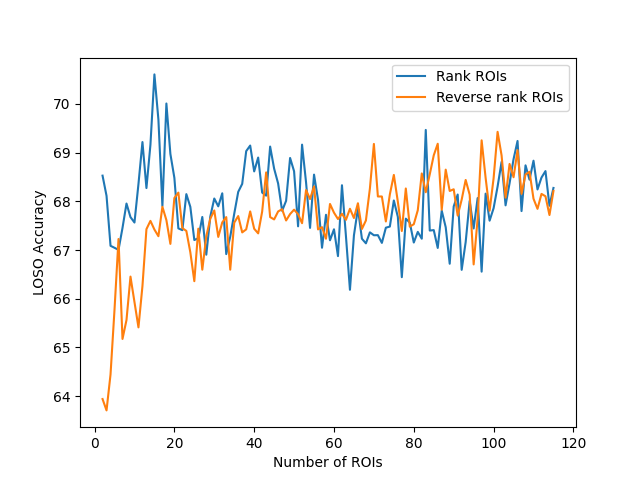}
  \caption{The accuracy as we increase the number of learning ROIs when we trained the SCCNN-RNN. (blue line) The accuracy in the case of learning by selecting in order of ROIs with the highest Rank. (orange line) The accuracy in the case of learning by selecting in order of ROI with the lowest Rank}
  \label{fig:fig5}
\end{figure}

 Fig \ref{fig:fig5} shows the result of the model’s accuracy trained with ranking ROIs. From the (blue line) on Fig \ref{fig:fig5}, two facts can be observed. First, it can be confirmed that high accuracy is obtained when learning by selecting ROIs in a certain order is performed, where the learning is done by selecting ROIs in the order obtained. It can be seen that some ROIs play an important role for the diagnosis. As shown in Fig \ref{fig:fig5}, the ranking that achieves high accuracy is the case of using the ROI in the top 20. In particular, an accuracy of 70.6\% was obtained when ROI up to 15th Rank was used for training. Another fact is that as the ROI used for learning increases, it is hard to tell areas that provide higher results than before. This means that using lots of ROIs can be a hindrance in diagnosing ADHD. When selecting the ROI in reverse order, we can see the results with the (orange line) on Fig \ref{fig:fig5}. Relatively low accuracy is obtained when using up to ~20th rank of reverse ranks, but it can be seen that the accuracy tends to be increased when gradually more ROIs is used. In other words, It means that learning with ROIs of low importance determined by accuracy does not help much in the diagnosis, but if the number of ROIs used for subsequent learning increases, it could be helpful. That is, ROIs with a high rank complement ROIs with a low rank.

%%%%%%%%%%%%%%%%% experiment section5 %%%%%%%%%%%%%%%%%%%%%%%
\subsection{Accuracy analysis for ADHD discrimination according to ROI ranking in  modified models}
\begin{figure}[t]
  \centering
  \includegraphics[scale=0.5]{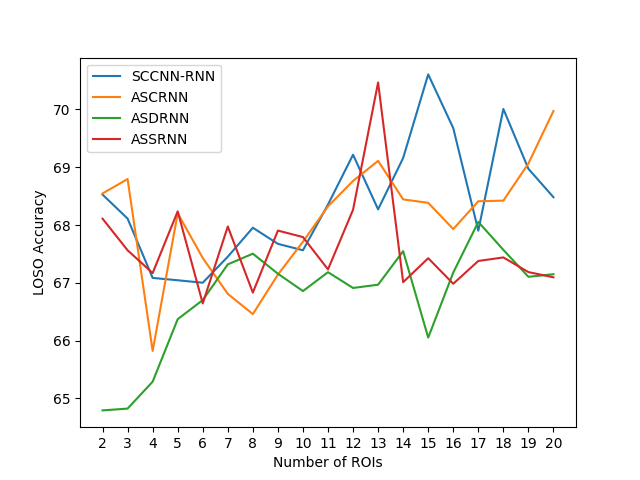}
  \caption{The accuracy when we trained the other modified models(ASCRNN, ASDRNN, ASSRNN) as we increase the number of  ROIs based the rank order}
  \label{fig:fig6}
\end{figure}

To investigate the impact on importance of ROI, we perform experiments with other modified models (ASCRNN, ASDRNN, ASSRNN) that we mention in section \ref{sec:method4}. The experiment is performed in the same way as described in section \ref{sec:experiment4}. However, the number of ROI used for learning was up to the 20th rank, as a result of Fig\ref{fig:fig5}. With Fig \ref{fig:fig6} and Table \ref{tab:table2}, we can describe the result as follows:First, when learned with selected ROIs according to its rank, higher results can be expected than using all ROIs. Further, all models provide accuracy around 68\% to 70\% despite using fewer areas than all 116 areas. In particular(Table \ref{tab:table2}), in the case of SCCNN-RNN and ASSRNN, the accuracies of 70.6(15)\% and 70.46(13)\% can be obtained, which are exceeding 70\% accuracy. This is a result that exceeds the previous result (68.6\%) by 2\%. Second (Fig \ref{fig:fig6}), with all the modified models based on SCCNN-RNN, the number of ROIs providing good accuracy lies between 10 and 20. This is slightly different from the case where only a small number of areas (ex. 1-5 rank ROI) are used. To summarize, it seems that a specific part of the brain must be used to improve the diagnosis. The best accuracy was obtained for the simplest model(SCCNN - RNN). There is an improvement of about 7\%, in the case of using ROIs with higher rankings. 

\begin{table}
  \centering
  \begin{tabular}{cccc}
    \toprule
    \multirow{2}{*}{Proposed paper}    & \multirow{2}{*}{Model name}     & \multirow{2}{*}{LOSO accuracy with all ROls} & LOSO accuracy with specific ROIs \\
    &&&(Number of ROIs used in the experiment) \\
    \midrule
\cite{riaz2017fcnet}  &FCNet& 60.4& - \\
\cite{riaz2018deep}  &DeepfMRI& 67.9 & - \\
\cite{RN7}  &SCCNN - Attention& 68.6 & - \\
\cite{RN7}  &SCCNN - RNN& 63.6 & 70.6(15) \\
    \midrule
- &ASCRNN& 65.2& 69.97(20) \\
- &ASDRNN& 68.4& 68.05(17) \\
- &ASSRNN& 66.86& 70.46(13) \\
    \bottomrule
  \end{tabular}
   \caption{Comparison of accuracy in ADHD-200 classification models using AAL template}
  \label{tab:table2}
\end{table}

%%%%%%%%%%%%%%%%% Conclusion %%%%%%%%%%%%%%%%%%%%%%%
\section{Conclusion}
\label{sec:conclusion}
Recently, the diagnosis accuracy of ADHD and HC has been gradually improved with rs-fMRI data through deep learning methods. However, there is difficulty improving the performance of deep learning. Even though there are many reasons, one of them is the limitation of medical data related to ADHD. Also, it is hard to find the medical biomarker that distinguishes between ADHD and normal. Nevertheless, rs-fMRI data are frequently used because they contain various information about the brain. In this study, we showed that finding ROI providing important areas in the brain can help diagnose ADHD. Moreover, it was found that only using 15  ROIs in the importance order can provide a more outstanding performance improvement (70.6\%) in the diagnosis than using all areas. It implies that establishing the good criteria for the importance of the ROI can give diagnostic accuracy. Thus, we will consider a critical area detection method using deep neural networks in the future study.
Further, in deep learning models, their decision process seems to be a black box. However, this can be a major drawback in diagnosing diseases. Therefore, we will look at how deep learning models can understand ADHD by studying the decision process of various deep learning models.

\section*{Acknowledgments}
This work is supported by the Basic Science Research Program through the National Research Foundation
of Korea funded by the Ministry of Education, Science and Technology (NRF2018R1D1A1B07049420) and Institute of Information \& communications Technology Planning \& Evaluation(IITP) grant funded by the Korea government(MSIT) (No.2020-0-01343, Artificial Intelligence Convergence Research Center (Hanyang University ERICA).

%Bibliography
%\begin{filecontents}{\references.bib}
%\end{filecontents}

\bibliographystyle{abbrvnat}
\bibliography{references}

\end{document}